\documentclass[aps,prc,preprint,showpacs,groupedaddress]{revtex4}
\begin{document}
\title{Decay Rate of Triaxially-Deformed Proton Emitters}

\author{Cary~N.~Davids and Henning Esbensen}

\affiliation{Physics Division, Argonne National Laboratory, Argonne, IL 60439}

\date{\today} 
\begin{abstract}
The decay rate of a triaxially-deformed proton emitter is calculated 
in a particle-rotor model, which is based on a deformed Woods-Saxon 
potential and includes a deformed spin-orbit interaction.  The wave 
function of the $I=7/2^{-}$ ground state of the deformed proton 
emitter $^{141}$Ho is obtained in the adiabatic limit, and a Green's 
function technique is used to calculate the decay rate and branching 
ratio to the first excited 2$^{+}$ state of the daughter nucleus.  
Only for values of the triaxial angle $\gamma$ $<5^{\circ}$ is good 
agreement obtained for both the total decay rate and the 2$^{+}$ 
branching ratio.
\end{abstract} 
\maketitle
\section{Introduction}
The fundamental simplicity of the proton decay process in nuclides 
whose ground states are unstable to proton emission has enabled a good 
deal of nuclear structure information to be obtained on nuclei beyond 
the proton drip line \cite{woods}.  The observable quantities are the 
proton energies and half-lives.  In the rare-earth region, the proton 
emitters are predicted to have large static quadrupole deformations 
\cite{moller}.  For these cases, analysis of the measurements has been 
carried out using a particle-rotor model, with the unbound proton 
interacting with an axially-symmetric deformed core 
\cite{bugrov,kadmensky,ferreira,maglione1,maglione2,sonzogni,kruppa,barmore,esbensen}. 
 The results of such analyses over the past several years has been to 
obtain a good description of the ground-state properties of deformed 
rare-earth proton emitters, including deformations, occupation 
factors, Nilsson orbitals for the decaying protons, and wave function 
decompositions.

Adding to the information provided by the observation of decay 
protons, recent measurements have been made of the level structure of 
the deformed proton emitter $^{141}$Ho by means of in-beam 
$\gamma$-ray spectroscopy \cite{seweryn}.  Particle-rotor calculations 
of the energy levels in the rotational band lying above the 
$I=7/2^{-}$ ground state suggest that better agreement with experiment 
would be obtained if the nuclear shape possessed a small amount of 
static triaxial deformation \cite{seweryn}.  From the standpoint of 
proton radioactivity it is therefore of interest to investigate the 
effect of a static triaxial deformation on the decay rate of a 
deformed proton emitter.  In this work we present such an analysis, 
and obtain numerical results for the decay rate of the deformed proton 
emitter $^{141g}$Ho.

\section{Coupled Equations in the $R$ Representation}
We generalize the treatment of Esbensen and Davids \cite{esbensen}, 
extending it to include the case of nuclei without axial symmetry.  
Using Eq.  5A-2 of ref.  \cite{BM-II}, we write the wave function of 
an odd-A even-N nucleus consisting of a proton coupled to an even-even 
triaxially-deformed rotor, in the laboratory (space-fixed) system as 
\begin {equation}\Psi_{IM}({\bf 
r},\omega)=\sum_{ljR\tau}\frac{\phi^{I}_{ljR\tau}(r)}{r}|ljR\tau 
IM\rangle,\label{wf} \end{equation} where $l$ and $j$ are the orbital and total 
angular momentum of the particle, $R$ and $\tau$ are the rotational 
quantum numbers of the rotor, and $I$ is the total angular momentum of 
the nucleus ({\bf I} = {\bf j} + {\bf R}).  The ket $|ljR\tau 
IM\rangle$ describes the dependence on spin and angular coordinates of 
the particle and the orientation angle of the rotor, and is given by 
\begin{equation}|ljR\tau IM\rangle=\sum_{mM_{R}}\langle 
jmRM_{R}|IM\rangle |R\tau M_{R}\rangle 
|ljm\rangle.\label{ketR}\end{equation} This is the laboratory frame or 
R-representation as described in Eq.  (2) of ref.  \cite{esbensen}.  
The total Hamiltonian of the proton-core system, 
\begin{equation}H=T+V({\bf r},\omega)+V_{ls}({\bf 
r},\omega)+H_{R},\label{ham}\end{equation} consists of the relative kinetic 
energy $T$, the nuclear plus Coulomb interaction $V({\bf r},\omega)$, 
which depends on the position ${\bf r}$ of the proton and the 
orientation $\omega$ of the rotor in the space-fixed system, the 
deformed spin-orbit potential $V_{ls}({\bf r},\omega)$, and the 
Hamiltonian $H_{R}$ of the rotor.  The detailed parametrization of the 
nuclear and Coulomb interactions are given in Appendix A, and the 
deformed spin-orbit term is discussed in Appendix B. To proceed, we 
first expand the potential $V({\bf r},\omega)$ in the D-functions, 
which are related to the spherical harmonics 
\begin{mathletters}\begin{eqnarray}V({\bf r},\omega)&= &\sum_{\lambda 
\mu}V_{\lambda\mu}(r)D^{\lambda}_{\mu 0}(\theta',\phi'), \\
V_{\lambda\mu}(r)& =&\frac{2\lambda+1}{4\pi}\int 
_{-1}^{1}d(cos\theta')\int_{0}^{2\pi} 
d\phi'D^{\lambda*}_{\mu0}(\theta',\phi')V(r,\theta',\phi'), 
\end{eqnarray}\end{mathletters}where $\theta^{'},\phi^{'}$ refer to 
the angles of the particle with respect to the 3-axis of the rotor.  
Because of reflection symmetry, $\lambda$ and $\mu$ are restricted to 
even values (see Appendix A).  Projecting with $|ljR\tau IM\rangle$ on 
the Schr\"{o}dinger equation $H\Psi_{IM}=E\Psi_{IM}$ we obtain a set 
of coupled equations in the radial wave functions 
\begin{eqnarray}(h_{lj}+E_{R\tau}-E)\phi^{I}_{ljR\tau}(r)&=& 
-\sum_{l'j'R'\tau'}\sum_{\lambda> 0, \mu}\langle ljR\tau 
IM|D^{\lambda}_{\mu 0} (\theta',\phi')|l'j'R'\tau'IM\rangle 
V_{\lambda\mu}(r) \phi^{I}_{l'j'R'\tau'}(r)\nonumber\ \\
&&-\sum_{l'j'R'\tau'}\langle ljR\tau IM|V_{ls}({\bf 
r},\omega)|l'j'R'\tau' IM\rangle \phi^{I}_{l'j'R'\tau'}(r) 
,\label{Ycoup} \end{eqnarray}where 
\[h_{lj}=\frac{\hbar^{2}}{2\mu_{0}}\left(-\frac{d^{2}}{dr^{2}}+ 
\frac{l(l+1)}{r^{2}}\right ) + V_{0}(r),\] and $V_{0}(r)$ is the 
monopole part of the Coulomb plus nuclear potential.  Here $\mu_{0}$ 
is the proton reduced mass, and $E_{R\tau}$ is the energy of the 
rotational state $|R\tau M_{R}\rangle$.  In Appendix B we will extract 
the monopole part of the spin-orbit potential for inclusion in 
$h_{lj}$.
\section{The K Representation}
The matrix elements on the right hand side of Eq.  (\ref{Ycoup}) are 
easiest to evaluate if we go over into the K-representation of ref.  
\cite{esbensen}, which is oriented in the body-centered coordinates of 
the rotor.  In this system the quantum number $\tau$ is identified 
with the projection $K_{R}$ of $R$ on the rotor's 3-axis.  For 
clarity, the quantum numbers $m$, $M_{R}$, and $M$ in Eq.  
(\ref{ketR}) all refer to angular momentum projections on the z-axis 
of the space-fixed coordinate system due to the particle, rotor, 
3-axis in the body-fixed system are denoted by $\Omega$, $K_{R}$, and 
$K$ (see Fig.  (\ref{vect})).  As a consequence, the following 
relations hold:
\begin{mathletters}
\begin{eqnarray}M=M_{R}+m \\ 
K=K_{R}+\Omega.\label{mbody}\end{eqnarray}\end{mathletters}

In the K-representation we can write the transformed rotor wave 
function as (see Eq.  4-7 of ref.\cite{BM-II})
\begin{equation} \langle\omega|R\tau M_{R}\rangle=\langle\omega|RK_{R}
M_{R}\rangle= 
\sqrt{\frac{2R+1}{8\pi^{2}}}D^{R}_{M_{R}K_{R}}(\omega),\label{rb} 
\end{equation}which is a function of the orientation $\omega$ of 
the rotor in the laboratory frame.  For an axially-symmetric rotor, we 
would have $K_{R}=0$, resulting in $\Omega=K$ from Eq.  (\ref{mbody}).  
The particle wave function is \begin{equation} 
|ljm\rangle=\sum_{\Omega}D^{j}_{m \Omega}(\omega)|lj\Omega 
\rangle,\label{pb}\end {equation}where the single-particle wave 
function $|lj\Omega \rangle$ is evaluated in the body-fixed frame of 
the rotor.

After inserting Eq.  (\ref{rb}) and (\ref{pb}) into Eq.  (\ref{ketR}), 
we may contract it using Eq.  (1A-43) of \cite{BM-II}, with the 
result\begin{equation}|ljRK_{R}IM\rangle= \sqrt{\frac{2R+1}{8\pi^{2}}} 
\sum_{K,\Omega} \langle j\Omega RK_{R}|IK\rangle 
D^{I}_{MK}(\omega)|lj\Omega\rangle.\label{unsym}\end{equation}

Because the rotor possesses symmetry after rotating 180$^{\circ}$ 
around any of its three axes, it is convenient to have the projection 
$K$ appear only as a positive number.  Symmetry properties of the wave 
function require that the quantity $K_{R}=K-\Omega$ be an even 
integer ($0,\pm 2,\pm 4\dots$) \cite{bohr}.  We then have:
\begin{equation}|ljRK_{R}IM\rangle=\sqrt{\frac{2R+1} 
{8\pi^{2}}} \sum_{K>0,\Omega}\langle j\Omega RK_{R}|IK\rangle \left[ 
D^{I}_{MK}(\omega)|lj\Omega \rangle+(-1)^{I-j} 
D^{I}_{M-K}(\omega)|lj\overline{\Omega} 
\rangle\right],\label{tria}\end{equation}where $\overline{\Omega}$ 
stands for $-\Omega$.  Since we are only interested in the low-lying 
states of the rotor where $R=0,2,4 \ldots $, we rewrite Eq.  
(\ref{tria}) as\begin{equation} 
|ljRK_{R}IM\rangle=\sum_{K>0,\Omega}A^{IK}_{j\Omega,RK_{R}}|lj\Omega 
KIM\rangle,\label{kexp}\end{equation} where\begin{equation} 
A^{IK}_{j\Omega,RK_{R}}=\sqrt{\frac{2R+1}{2I+1}}\langle j\Omega 
RK_{R}|IK\rangle \sqrt{1+(-1)^{R}}\label{ED14}\end{equation}and
\begin{equation}|lj\Omega KIM\rangle=\sqrt{\frac{2I+1}{16\pi^{2}}} 
\left[ D^{I}_{MK}(\omega)|lj\Omega \rangle +(-1)^{I-j} 
D^{I}_{M-K}(\omega)|lj \overline{\Omega} \rangle \right].\label{ED15} 
\end{equation}For an axially-symmetric nucleus, $K_{R}$ vanishes, 
making $\Omega=K$, and Eq. (\ref{ED14}) and (\ref{ED15}) become 
identical to Eq. (14) and (15) of ref. \cite{esbensen}.

Inserting Eq.  (\ref{kexp}) into Eq.  (1) we can now express the total 
wave function in terms of the new basis 
(\ref{ED15}):\begin{equation}\Psi_{IM}= 
\sum_{lj}\sum_{K>0}\sum_{\Omega}\frac{\phi^{IK}_{lj\Omega}(r)}{r}|lj\Omega 
KIM\rangle \label{wavK},\end{equation}where the radial wave functions 
are\begin{equation}\phi^{IK}_{lj\Omega}(r)= 
\sum_{RK_{R}}A^{IK}_{j\Omega,RK_{R}}\phi^{I}_{ljRK_{R}}(r) 
\label{radK}.\end{equation}Note that the triaxial radial 
wave functions depend on the particle quantum number $\Omega$ in 
addition to $lj$.

It is easy to show that the amplitudes (\ref{ED14}) form an 
orthonormal transformation between the K and the R representation, 
i.e.  \begin{equation}\sum_{K>0}\sum_{\Omega}A^{IK}_{j\Omega,RK_{R}} 
A^{IK}_{j\Omega,R'K_{R'}}= \delta_{R,R'}\delta_{K_{R},K_{R'}},\ \ \ \ 
\ \ \ \sum_{RK_{R}}A^{IK}_{j\Omega,RK_{R}}A^{IK'}_{j\Omega',RK_{R}}= 
\delta_{\Omega,\Omega'}\delta_{K,K'}\label{orth}.\end{equation}Thus we 
can transform the results obtained in one representation into the 
other.  After inverting Eq.  (\ref{radK}) we obtain 
\begin{equation}\phi^{I}_{ljRK_{R}}(r)= 
\sum_{K>0}\sum_{\Omega}A^{IK}_{j\Omega,RK_{R}} 
\phi^{IK}_{lj\Omega}(r).\label{radR}\end{equation}

\subsection{Coupled Equations in the $K$ Representation}

We continue with the evaluation of the matrix elements on the RHS of 
Eq.  (\ref{Ycoup}).  Inserting expression (\ref{kexp}) for the 
spin-angular wave functions in the R-representation we 
obtain\begin{eqnarray}\lefteqn{\langle ljRK_{R}IM|D^{\lambda}_{\mu 0} 
(\theta',\phi')|l'j'R'K_{R'}IM\rangle = }\ \ \ \ \ \ \nonumber \\
& & \sum_{ K>0}\sum_{K'>0}\sum_{\Omega,\Omega'} 
A^{IK}_{j\Omega,RK_{R}}\langle lj\Omega KIM|D^{\lambda}_{\mu 
0}(\theta^{'},\phi^{'})|l'j'\Omega' K'IM\rangle 
A^{IK'}_{j'\Omega',R'K_{R'}} \nonumber \end{eqnarray}and a similar 
expression for the matrix element of the spin-orbit potential.  The 
advantage of using the K representation, Eq.  (\ref{ED15}), now 
becomes evident because each matrix element is the product of two 
parts.  The first part involves integrating a product of orthogonal 
D-functions over the orientation coordinates $\omega$ of the rotor, 
and yields the important result $\delta_{K,K'}$.  The second part 
involves the single-particle wave functions in the body-fixed rest 
frame of the rotor: \begin{equation}\langle lj\Omega|D^{\lambda}_{\mu 
0}(\theta',\phi')|l'j'\Omega' \rangle =(-1)^{\lambda} \langle 
j\mbox{$\frac{1}{2}$}\lambda 0|j'\mbox{$\frac{1}{2}$}\rangle \langle 
j'\Omega'\lambda\mu|j\Omega\rangle.\label{matri}\end{equation}Thus
\begin{equation}\langle ljRK_{R}IM|D^{\lambda}_{\mu 0} 
(\theta',\phi')|l'j'R'K_{R'}IM\rangle 
=\sum_{K'>0}\sum_{\Omega,\Omega'} \!  A^{IK'}_{j\Omega,RK_{R}}\langle 
lj\Omega |D^{\lambda}_{\mu 0}(\theta',\phi')|l'j'\Omega'\rangle 
A^{IK'}_{j'\Omega',R'K_{R'}}.\label{ylm}\end{equation}For the 
spin-orbit potential we have a similar expression.  We show how to 
evaluate the matrix elements of the spin-orbit interaction in Appendix 
B. Because the matrix elements are diagonal in $K$, Eq.  (\ref{mbody}) 
shows that the particle projections $\Omega,\Omega'$ are restricted by 
$|\Omega-\Omega'|$ = an even integer.  We still have the previously 
determined restrictions $\lambda$ even and $|K-\Omega|$ = an even 
integer.

We now obtain the coupled equations in the $K$ representation by 
multiplying Eq.  (\ref{Ycoup}) by $A^{IK}_{j\Omega,RK_{R}}$ and 
summing over $RK_{R}$, using Eq.  (\ref{radK}) and (\ref{orth}).  The 
presence of the rotational energy $E_{RK_{R}}$ requires the use of Eq.  
(\ref{radR}) for the radial wave function.  Thus we obtain

\begin{eqnarray}\lefteqn{(h_{lj}-E)\phi^{IK}_{lj\Omega}(r)+ 
\sum_{K'>0}\sum_{\Omega'}W_{j\Omega\Omega'}^{KK'} 
\phi^{IK'}_{lj\Omega'}(r)=} \ \ \ \ \ \ \ \nonumber \\
& & -\sum_{l'j'}\sum_{\Omega'}\left\{ \sum_{\lambda>0, \mu}\langle 
lj\Omega|D^{\lambda}_{\mu 0}(\theta',\phi')|l'j'\Omega'\rangle 
V_{\lambda\mu}(r) +\langle lj\Omega|V_{ls}({\bf r})|l'j'\Omega'\rangle 
\right\} \phi^{IK}_{l'j'\Omega'}(r),\label{coup}\end{eqnarray} where 
\begin{equation}W_{j\Omega\Omega'}^{KK'}= 
\sum_{RK_{R}}A^{IK}_{j\Omega,RK_{R}} 
E_{RK_{R}}A^{IK'}_{j\Omega',RK_{R}}.\label{Wkk}\end{equation}To recover the 
axially-symmetric case we set $\Omega=\Omega'=K$, and $\mu=0$ in Eq.  
(\ref{coup}) and (\ref{Wkk}).  This removes the sums over $\Omega'$, 
$K_{R}$, and $\mu$.

\subsection{Adiabatic limit}
For a comparison with the results obtained in \cite{esbensen}, we will 
solve the coupled equations (\ref{coup}) in the adiabatic limit, where 
the rotational energies $E_{RK_{R}}$ of the core are set to zero.  
This sets equal to zero the second term on the left-hand side of Eq.  
(\ref{coup}).  In this case we have 
\begin{equation}(h_{lj}-E)\phi^{IK}_{lj\Omega}(r)= 
-\!\sum_{l'j'\Omega'}\left\{ \sum_{\lambda>0, \mu}\langle 
lj\Omega|D^{\lambda}_{\mu 0}(\theta',\phi')|l'j'\Omega'\rangle 
V_{\lambda\mu}(r) +\langle lj\Omega|V_{ls}({\bf r})|l'j'\Omega'\rangle 
\right\} \phi^{IK}_{l'j'\Omega'}(r).\label{coupad}\end{equation}The 
important thing to notice is that the coupled equations are then 
diagonal in $K$, but include $\Omega$-mixing.  In other words, in the 
adiabatic limit, $K$ is still a good quantum number, but, in addition 
to the $\Omega=K$ component, the interaction mixes into the wave 
function components with $\Omega=K\pm 2,K\pm 4$, etc, subject to the 
restriction $|\Omega|\leq j$.  It should also be noted that the 
Clebsch-Gordan coefficient in Eq.  (\ref{matri}) is non-zero only when 
$\Omega'+\mu=\Omega$.

The consequence of this $\Omega$-mixing is that the number of coupled 
equations will be substantially larger than is found in the 
axially-symmetric case.  As an example, consider the combinations of 
$j$ and $\Omega$ needed to solve the equations for the 
$I=K=\frac{7}{2}^{-}$ ground state of the deformed proton emitter 
$^{141}$Ho.  The spherical states involved will be 
$j=\frac{7}{2}^{-}$, $\frac{9}{2}^{-}$, $\frac{11}{2}^{-}$, 
$\frac{13}{2}^{-}$, and $\frac{15}{2}^{-}$.  In addition to 
$\Omega=K$, for each $j$ there will be associated up to 7 more values 
of $\Omega$.  Table \ref{jomega} shows the permissible $j,\Omega$ 
combinations for the $I=K=\frac{7}{2}^{-}$ ground state of $^{141}$Ho.  
The total number of wave function combinations is 30, which is to be 
contrasted with only 5 for the axially-symmetric case.

\subsection{Decay Rate Calculation}

We obtain the partial decay rate for proton emission from a state 
having angular momentum $I=K$ in the adiabatic limit via either the 
direct method (Dir) or the distorted wave Green's function method (DW) 
using Eq.  (7) of \cite{esbensen}.  While in the axially-symmetric 
case the daughter states were labeled only by the quantum number $R$, 
in the triaxial case we need the second label $\tau$ or $K_{R}$ as 
well: \begin{equation}\Gamma^{I}_{RK_{R}}= 
\sum_{lj}\Gamma^{I}_{ljRK_{R}}= 
\frac{\hbar^{2}k_{R}}{\mu}\sum_{lj}|N^{I,Dir/DW}_{ljRK_{R}}|^{2} 
,\label{ndir} \end{equation}where\begin{eqnarray}N^{I,Dir}_{ljRK_{R}} 
& = & \frac{\phi^{I}_{ljRK_{R}}(r)}{G_{l}(k_{R}r)}\ \ \ \mbox{at $r = 
r_{m}$,} \nonumber \\N^{I,DW}_{ljRK_{R}} & = & - 
\frac{2\mu}{\hbar^{2}k_{R}}\sum_{l'j'R'K_{R'}} \int_{0}^{r_{int}}dr 
F_{l}(k_{R}r) \langle ljRK_{R}IM|V({\bf r},\omega)+V_{ls}({\bf 
r},\omega)\nonumber \\&&-\frac{Z_{D}e^{2}}{r}|l'j'R'K_{R'}IM\rangle 
\phi^{I}_{l'j'R'K_{R'}}(r).  \label{ndw1}\end{eqnarray}Here $k_{R}$ is 
the proton wave number for the decay to the daughter state with 
quantum numbers $RK_{R}$.  Having obtained the radial wave function 
$\phi^{IK}_{lj\Omega}(r)$ for a given $K=I$ in the adiabatic limit, we 
can construct the associated radial wave functions in the R 
representation\begin{equation} \phi^{I}_{ljRK_{R}}(r)= 
\sum_{\Omega}A_{j\Omega,RK_{R}}^{IK}\phi^{IK}_{lj\Omega}(r)\end{equation} 
by means of Eq.  (\ref{radR}).  Inserting this wave function into Eq.  
(\ref{ndw1}), we obtain for the Direct 
method\begin{equation}N^{I,Dir}_{ljRK_{R}}=\frac{1}{G_{l}(k_{R}r_{m})} 
\sqrt{\frac{2(2R+1)}{2I+1}} \sum_{\Omega}\langle j\Omega 
RK_{R}|IK\rangle\phi^{IK}_{lj\Omega}(r_{m}).\end{equation}Using this 
equation we can immediately see the effect of the triaxial shape on 
the decay rate: for decay to the ground state ($R=0$), the 
Clebsch-Gordan coefficient has the value $\delta_{j,I}\delta_{\Omega, 
K}$.  Thus only the $\Omega=K$ component of the wave function will 
participate in this decay branch, effectively reducing its decay rate 
relative to the that for the axially-symmetric case since there are 
now other $\Omega$-components present in the triaxial wave function.

For the Green's function method, after performing the same operations 
that led to Eq. (\ref{ylm}) to obtain the matrix elements of the 
interaction in Eq.  
(\ref{ndw1}), the second orthogonality relation in Eq.  (\ref{orth}) 
allows the summations over $R'$ and $K_{R'}$ to be performed.  The 
final expression for the Green's function method is 
\begin{eqnarray}N^{I,DW}_{ljRK_{R}} & = & - 
\frac{2\mu}{\hbar^{2}k_{R}}\sum_{l'j'}\sum_{\Omega,\Omega'} 
A^{IK}_{j\Omega,RK_{R}} \int_{0}^{r_{int}}dr F_{l}(k_{R}r)\langle 
lj\Omega|V({\bf r},\omega)+V_{ls}({\bf r},\omega) \nonumber \\& 
&-\frac{Z_{D}e^{2}}{r}|l'j'\Omega'\rangle\phi^{IK}_{l'j'\Omega'}(r).  
\label{ndw2}\end{eqnarray}This expression reduces to the analogous one 
for an axially-symmetric nucleus when $K$ is set equal to $\Omega$ and 
the sums over $\mu$, $\Omega$ and $\Omega'$ are removed.
\section{Applications}

In this section we use the formalism described above to calculate 
decay rates for the deformed proton emitter $^{141}$Ho 
($I=K=7/2^{-})$, in both the axially-symmetric and triaxial cases.  
The calculations have been performed in the adiabatic limit, using the 
potential parameters found in ref.  \cite{esbensen}.  The ground-state 
wave function was expanded in spherical components with $j=7/2^{-}$, 
$9/2^{-}$, $11/2^{-}$, $13/2^{-}$, and $15/2^{-}$.  All multipole 
expansions were carried out up to $\lambda=14$.  The results are 
compared with experimental values for the total decay width and the 
branching ratio for decay to the first excited 2$^{+}$ state in the 
daughter nucleus $^{140}$Dy.

\subsection{Axially-Symmetric Case: the Deformed Proton Emitter $^{141}$Ho}

To verify the formulation, we have computed the decay half-life and 
branching ratio to the 2$^{+}$ state in the daughter nucleus for the 
deformed proton emitter $^{141}$Ho , using deformed spin-orbit matrix 
elements for the axially-symmetric case ($\gamma=\mu=0,\ 
\Omega=\Omega'=K$) calculated from Eq.  (\ref{matel}), which, again 
for $\lambda>0$, now reads\begin{equation}\langle 
ljK|V_{ls}(r,\theta)|l'j'K\rangle 
=2V_{so}\sum_{\lambda>0}\frac{\langle j'K\lambda 
0|jK\rangle}{\sqrt{2j+1}}\langle lj\|\left[\nabla 
f_{\lambda\mu}(r)D^{\lambda}_{\mu0}({\bf\hat{r}})\right] \!\cdot 
(-i\nabla\times\hat{\sigma}\|l'j'\rangle,\end{equation} where the 
reduced matrix element is that calculated in Eq.  (\ref{RMA}).  The 
results of the calculations are shown in Table \ref{141}, and are 
identical with those obtained using the axially-symmetric deformed 
spin-orbit term of ref.  \cite{esbensen}.  It should be noted that the 
numbers obtained here and in ref.  \cite{esbensen} differ slightly, 
since the $2^{+}$ excitation energy of 202 keV in the daughter nuclide 
$^{140}$Dy used here has only recently been measured 
\cite{cullen,krolas}, and was not known when the calculations in ref.  
\cite{esbensen} were carried out.  In that work an excitation energy of 
166 keV was used.
  
\subsection{Triaxial Case: the Deformed Proton Emitter 
$^{141}$Ho} 

As previously mentioned, particle-rotor calculations of the energy 
levels in the rotational band lying above the $I=7/2^{-}$ ground state 
of the deformed proton emitter $^{141}$Ho were reported in ref.  
\cite{seweryn}.  These calculations suggest that better agreement with 
experiment would be obtained with the introduction of a small amount 
of static triaxial deformation.  In order to assess the effect on the 
proton decay rates of adding such a deformation, we have used the 
formalism developed in the present work to compute, in the adiabatic 
limit, the decay rate and 2$^{+}$ branching ratio for this proton 
emitting nuclide, as a function of the angle of triaxiality $\gamma$.  
The calculations were performed with $\beta_{2}=0.244$, 
$\beta_{4}=-0.046$.  Preliminary results have been reported in ref.  
\cite{davids2}.

Fig.  \ref{width} shows the product of calculated total decay width 
$\Gamma_{calc}$ and spectroscopic factor $S_{calc}=u^{2}=0.6$ for the 
$^{141}$Ho ground state, plotted as a function of the triaxial angle 
$\gamma$.  The spectroscopic factor was obtained from a BCS 
calculation, using a proton pairing gap $\Delta_{p}$ of 0.9 MeV. The 
shaded area represents the experimental measurement, and the error bar 
attached to the calculated curve represents the uncertainty in the 
calculated width due to the uncertainty in the proton energy.  It is 
seen that calculation and experiment agree well for small values of 
$\gamma$.

Fig.  \ref{bratio} shows the calculated 2$^{+}$ branching ratio, 
plotted as a function of $\gamma$, along with the experimental value, 
0.0070(15) \cite{ryka}.  The small error bar on the calculated curve 
is due to the uncertainty in the proton energy.  The agreement between 
calculation and experiment is excellent for small values of $\gamma$, 
and this suggests that a static triaxial deformation, if present, is 
limited to an angle of $\leq$ 5$^{\circ}$.

Plotted in Fig.  \ref{cj} are the amplitudes for the allowed 
$\Omega$-values of $+7/2^{-}$, $+3/2^{-}$, $-1/2^{-}$, and $-5/2^{-}$ 
for the $j=7/2$ spherical component of the $^{141}$Ho ground state 
wave function.  As expected, for $\gamma=0$, only the $\Omega=+7/2^{-}$ 
component is present, but with increasing $\gamma$, other 
$\Omega$-components differing by $\pm$2, $\pm$4 $\ldots$ begin to mix 
into the wave function.  The decay proceeds primarily to the daughter 
ground state, and only the $\Omega=+7/2^{-}$ component participates in 
this branch.  Thus the decrease in the total decay width with 
increasing $\gamma$ seen in Fig.  \ref{width} tracks with the decrease 
in the $\Omega=+7/2^{-}$ wave function component, which itself follows 
from the increasing appearance of other $\Omega$-values in the wave 
function.

\section{Conclusions} In this work we have developed a formalism to 
include the effect of static triaxial deformation on calculations of 
the decay rate for a deformed nucleus.  The main complications over 
the axially-symmetric case are the additional dependence of the 
interaction potential on the proton azimuthal angle $\phi$, and a 
consequent increase in the complexity of the deformed spin-orbit 
potential.  The extra dimension causes additional $\Omega$-components 
to be introduced into the wave function.  The matrix elements of the 
spin-orbit interaction have been calculated using a tensor algebra 
approach, as described in Appendix B.

We have applied the abovementioned methodology to calculations of the 
decay rate of the deformed proton emitter $^{141g}$Ho, which has 
spin-parity $7/2^{-}$.  In ref.  \cite{esbensen} it was shown that the 
non-adiabatic coupled-channels approach does not yield results for 
$^{141g}$Ho decay which agree with experiment, either for the absolute 
decay rate or the branching ratio for decay to the first 2$^{+}$ of 
the daughter nucleus $^{140}$Dy.  This is because of the presence of 
Coriolis mixing, which is particularly strong for high spin states.  
Empirically it is known that it is necessary to quench the Coriolis 
mixing in order to obtain good agreement with data 
\cite{henriquez,fiorin}.  It was found in \cite{esbensen} that good 
agreement with experiment was obtained in the adiabatic limit, where 
the energies of the rotational states of the daughter nucleus are set 
to zero.  For this reason we have performed the triaxial calculations 
in the adiabatic limit.

After first checking the results of the calculation against an 
axially-symmetric code, triaxiality was introduced, ranging up to a 
$\gamma$ angle of 40$^{\circ}$.  While the sensitivity of the 
resulting total decay rate to triaxial angle $\gamma$ was not high, 
the calculation of the branching ratio for decay to the first 2$^{+}$ 
state of the daughter showed a strong dependence on $\gamma$, essentially 
ruling out angles greater than 5$^{\circ}$.  We believe that the 
branching ratio calculation is quite reliable, since factors such as 
absolute spectroscopic factors tend to cancel.

In conclusion, we do not believe that a static triaxial deformation 
plays an important role in helping to explain the decay rate of the 
deformed proton emitter $^{141g}$Ho.  Triaxiality may still exert an 
influence on the nuclear structure of this nuclide, but most likely at 
higher spins and excitation energies.

\begin{acknowledgments}
This work was supported by the U. S. Department of Energy, Nuclear 
Physics Division, under Contract No. W-31-109-ENG-38.
\end{acknowledgments}
\appendix
\section{Parametrization of the Interaction Potential}

For triaxial nuclei, the degree of triaxiality is denoted by the angle 
$\gamma$, which is one of the Hill-Wheeler \cite{hw} coordinates.  We 
parametrize the nuclear interaction between the valence proton and the 
deformed core nucleus in terms of the Fermi function 
$f(x)=[1+exp(x)]^{-1}$ as\begin{equation}V_{N}(r,\theta,\phi)= 
V_{N}^{(0)}f\left(\frac{r-R(\theta,\phi)}{a(\theta)}\right),\label{np} 
\end{equation}where $V_{N}^{(0)}$ is the depth of the nuclear 
potential, $a(\theta)$ is an angle-dependent diffuseness as in Eq.  
(A6) of \cite{esbensen}, and\begin{equation}R(\theta,\phi) = 
R_{N}\left\{ 
1+\sum_{\mu=-2,even}^{2}\!\!\!\!\!\!a_{2\mu}Y_{2\mu}(\theta,\phi)+ 
\sum_{\mu=-4,even}^{4}\!\!\!\!\!\!a_{4\mu}Y_{4\mu}(\theta,\phi)\right\}.
\label{radius}\end{equation}Here ($\theta,\phi)$ are the angles between {\bf r}
and the 3-axis of the core, and \[a_{20}= \beta_{2}\,\mbox{cos}\gamma\ 
\ \ \ \ a_{22}=a_{2-2}= \frac{\beta_{2}\, \mbox{sin}\gamma}{\sqrt{2}}\ 
\ \ \ \] 
\[a_{40}=\mbox{$\frac{1}{6}\beta_{4}\,$(5\,cos}^{2}\gamma+1)\] 
\[a_{42}=a_{4-2}=-\mbox{$\frac{1}{12}$}\beta_{4}\,\sqrt{30}\,\mbox{sin}2\gamma\] 
\begin{equation}a_{44}=a_{4-4}=\mbox{$\frac{1}{12}$}\beta_{4}\,\sqrt{70}\, 
\mbox{sin}^{2}\gamma.  \end{equation} The radius is calculated as 
\begin{equation}R_{N}=r_{0}\left ( 
\frac{A_{D}}{C(\beta_{2},\beta_{4},\gamma)}\right 
)^{\frac{1}{3}},\label{RN} 
\end{equation}where\begin{equation}C(\beta_{2},\beta_{4}) 
=\int\frac{d\Omega}{4\pi}\left ( 
1+\sum_{\mu=-2,even}^{2}\!\!\!\!\!\!a_{2\mu}Y_{2\mu}(\theta,\phi)+ 
\sum_{\mu=-4,even}^{4}\!\!\!\!\!\!a_{4\mu} Y_{4\mu}(\theta,\phi)\right 
)^{3} \label{vol}\end{equation}is the volume preserving factor and 
$A_{D}$ is the mass number of the core.  Equations (\ref{radius}) and 
(\ref{vol}) reduce to Eq.  (A1) and (A2) of \cite{esbensen} for the 
axially--symmetric case ($\mu=\gamma=0)$.  Values of $r_{0}=1.25$ fm and 
$a=0.65$ fm are used in our calculations.

We parametrize the charge density of the core in a similar way, using 
the slightly different radial parameters of $r_{0}=1.22$ fm and 
$a_{C}=0.56$ fm, as was done in ref. \cite{esbensen}.

\section{Deformed Spin-Orbit Potential}

For the solution of the coupled equations Eq.  (\ref{coupad}), matrix 
elements of the spin-orbit potential $\langle lj\Omega|V_{ls}(r, 
\theta,\phi)|l'j'\Omega'\rangle $ are needed.  As in ref.  
\cite{esbensen} we use a deformed spin-orbit potential, but with the 
addition of $\phi$-dependent terms not found there.  We follow the 
tensor algebra approach of Hagino \cite{hagino}, who has calculated 
matrix elements of the spin-orbit interaction in the context of 
particle-vibration coupling for spherical nuclei.

The starting point is to express the deformed spin-orbit interaction 
in the Thomas form \cite{esbensen} with the deformed Fermi function of 
Eq.  (\ref{np}): 
\begin{equation}V_{ls}(r,\theta,\phi)=4V_{so}\left(\left[\nabla 
f\left(\frac{r-R(\theta,\phi)}{a}\right)\right]{\bf \times p}{\bf 
\cdot s}\right).\end{equation}We now make a multipole expansion of the 
Fermi function\begin{eqnarray} 
f(r,\theta,\phi)&=&\sum_{\lambda\mu}f_{\lambda\mu}(r) 
D^{\lambda}_{\mu0}({\bf\hat{r}}),\label{exp}\\
f_{\lambda\mu}(r)&=&\frac{2\lambda+1}{4\pi}\int 
D^{\lambda*}_{\mu0} ({\bf\hat{r}})f(r,\theta,\phi) 
d(\mbox{cos}\theta)d\phi\label{coeff},\nonumber\\
\frac{df_{\lambda\mu}(r)}{dr}&=&\frac{2\lambda+1}{4\pi}\int 
D^{\lambda*}_{\mu0} ({\bf\hat{r}})\frac{df(r,\theta,\phi)}{dr} 
d(\mbox{cos}\theta)d\phi.\end{eqnarray} So
\begin{eqnarray}V_{ls}(r,\theta,\phi)&=&4V_{so} 
\left(\sum_{\lambda\mu}\left[\nabla 
f_{\lambda\mu}(r)D^{\lambda}_{\mu0}({\bf\hat{r}})\right]{\bf \times 
p}{\bf \cdot s}\right) \nonumber \\
&=&4V_{so}\left(\frac{1}{r}\frac{df_{00}(r)}{dr}D^{0}_{00}({\bf\hat{r}})({\bf 
r\times \!p)}{\bf \cdot s} +\sum_{\lambda>0,\mu}\left[\nabla 
f_{\lambda\mu}(r)D^{\lambda}_{\mu0}({\bf\hat{r}})\right]{\bf \times p}{\bf 
\cdot s}\right)\nonumber \\
&=&4V_{so}\left(\frac{1}{r}\frac{df_{00}(r)}{dr}{\bf l\cdot 
s}+\frac{1}{2}\sum_{\lambda>0,\mu}\left[\nabla 
f_{\lambda\mu}(r)D^{\lambda}_{\mu0}({\bf\hat{r}})\right] 
\cdot(-i\nabla\times\hat{\sigma})\right)\label{soc} \end{eqnarray} The 
first term on the RHS of Eq.  (\ref{soc}) is the monopole part of the 
spin-orbit potential, \begin{equation}V_{ls}^{0}(r) 
=4V_{so}\frac{1}{r}\frac{df_{00}(r)}{dr}{\bf l\cdot s},\end{equation} 
which can easily be incorporated into the monopole part of the 
Hamiltonian.  The second term,\begin{equation}\delta 
V_{ls}(r,\theta,\phi)=V_{ls}(r,\theta,\phi)-V_{ls}^{0}(r),\end{equation} 
can be decomposed into a sum of angular momentum tensors involving the 
spherical harmonics $Y_{\lambda\mu}({\bf\hat{r}})$, using the gradient 
formula Eq.  (5.9.17) of \cite{edmonds}:\begin{eqnarray}\left[\nabla 
f_{\lambda\mu}(r)D^{\lambda}_{\mu0}({\bf\hat{r}})\right]\!  
\cdot\!(-i\nabla\times\hat{\sigma})& 
=&-\frac{\sqrt{4\pi(\lambda+1)}}{2\lambda+1} 
\left(\frac{df_{\lambda\mu}(r)}{dr}- 
\frac{\lambda}{r}f_{\lambda\mu}(r)\right)\left[Y_{\lambda+1} 
(-i\nabla\times\hat{\sigma})\right]^{\lambda\mu}\nonumber\\
&&+\frac{\sqrt{ 4\pi\lambda}}{2\lambda+1} 
\left(\frac{df_{\lambda\mu}(r)}{dr}+ 
\frac{\lambda+1}{r}f_{\lambda\mu}(r)\right)\left[Y_{\lambda-1} 
(-i\nabla\times\hat{\sigma})\right]^{\lambda\mu}\nonumber\\
&=&\frac{\sqrt{4\pi}}{2\lambda+1}\left\{ 
\sqrt{\lambda}Q_{\lambda\mu}(r) T_{-}^{\lambda\mu}- 
\sqrt{\lambda+1}P_{\lambda\mu}(r)T_{+}^{\lambda\mu}\right\}, 
\label{oper}\end{eqnarray}where
\begin{eqnarray}T_{\pm}^{\lambda\mu}&=&\left[Y_{\lambda\pm1} 
(-i\nabla\times\hat{\sigma})\right]^{\lambda\mu},\label{tlm}\\
P_{\lambda\mu}(r)&=&\frac{df_{\lambda\mu}(r)}{dr}- 
\frac{\lambda}{r}f_{\lambda\mu}(r),\nonumber\ \mbox{   and}\\
Q_{\lambda\mu}(r)&=&\frac{df_{\lambda\mu}(r)}{dr}+ 
\frac{\lambda+1}{r}f_{\lambda\mu}(r).\nonumber\end{eqnarray} 

To solve the coupled radial equations we need to obtain the matrix 
elements of the operator $\sum_{\lambda>0,\mu}\left[\nabla 
f_{\lambda\mu}(r)D^{\lambda}_{\mu0}({\bf\hat{r}})\right] 
\cdot(-i\nabla\times\hat{\sigma})$, or specifically, the matrix 
elements of $\sum_{\lambda>0,\mu}T_{\pm}^{\lambda\mu}$ between 
spin-angular momentum states.  This is done by using the Wigner-Eckart 
theorem\begin{equation}\langle 
lj\Omega|T_{\pm}^{\lambda\mu}|l'j'\Omega'\rangle= \langle 
j'\Omega'\lambda\mu|j\Omega\rangle\frac{\langle 
lj\|T_{\pm}^{\lambda}\|l'j'\rangle}{\sqrt{2j+1}}\label{WE}\end{equation} 
along with reduced matrix elements given in Eq.  (58) and (59) of ref.  
\cite{reinhard} and Eq.  (A2.14) of ref.  \cite{lawson}.  Eq.  
(\ref{WE}) is the point of departure between axially-symmetric and 
triaxial calculations of the spin-orbit matrix elements, since the 
reduced matrix elements are independent of the projection quantum 
numbers $\Omega'$, $\Omega$, and $\mu$.  For axial symmetry we will 
set $\Omega=\Omega'=K$ and $\gamma=\mu=0$.

Putting $L_{\pm}=\lambda\pm 1$ we have\begin{eqnarray}\langle 
lj\|T_{\pm}^{\lambda}\|l'j'\rangle&=&\langle lj\|\left[Y_{L_{\pm}} 
(-i\nabla\times\hat{\sigma})\right]^{\lambda}\|l'j'\rangle \nonumber \\
&=& \sum_{l_{\alpha}j_{\alpha}}(-1)^{\lambda+j+j'+1} 
\sqrt{36(2\lambda+1)(2j'+1)(2j_{\alpha}+1)} \nonumber \\ 
&& \times\left \{\begin{array}{ccc}L_{\pm}&1&\lambda 
\\j'&j&j_{\alpha}\end{array}\right \}\left 
\{\begin{array}{ccc}l_{\alpha}&l'&1 \\
\frac{1}{2}&\frac{1}{2}&1 \\j_{\alpha}&j'&1 \end{array}\right \} 
\langle lj\|Y_{L_{\pm}}\|l_{\alpha}j_{\alpha}\rangle \langle 
Y_{l_{\alpha}}\|\nabla\|Y_{l'}\rangle, 
\label{mat}\end{eqnarray}where\begin{equation} \langle 
lj\|Y_{L_{\pm}}\|l_{\alpha}j_{\alpha}\rangle=(-1)^{\frac{1}{2}+j} 
\sqrt{\frac{(2j+1)(2L_{\pm}+1)(2j_{\alpha}+1)}{4\pi}} 
\left(\begin{array}{ccc}j&L_{\pm}&j_{\alpha} 
\\\frac{1}{2}&0&-\frac{1}{2}\end{array} 
\right)\end{equation}and\begin{equation}\langle 
Y_{l_{\alpha}}\|\nabla\|Y_{l'}\rangle= \sqrt{2l'+1}\langle 
l'010|l_{\alpha}0\rangle \left[ \frac{d}{dr}+\frac{1}{r} +\frac{1}{2r} 
\left\{ l'(l'+1)-l_{\alpha}(l_{\alpha}+1)\right\} \right].
\label{der}\end{equation} 
From the properties of the Clebsch-Gordan coefficient in Eq.  
(\ref{der}) it can be seen that $l_{\alpha}$ can take on only the 2 
values $l'\pm 1$, indicating that the parity of $l_{\alpha}j_{\alpha}$ 
is opposite to that of $l'j'$ and $lj$.  The permissible values of 
$j_{\alpha}$ in the summation in Eq.  (\ref{mat}) are 
$j_{\alpha}=l_{\alpha}\pm 1/2$, and may be further restricted by the 
triangle relations for the 6j-symbol, namely 
$\Delta(L_{\pm}jj_{\alpha})$ and $\Delta(j'1j_{\alpha})$.  
Additionally, for the calculation of the matrix elements, the $d/dr$ 
term in Eq.  (\ref{der}) needs to be modified, since the coupled 
equations are in the wave function $\phi_{l}(r)$, while the 
Hamiltonian (\ref{ham}) acts on the wave function (\ref{wf}), which 
contains $\phi_{l}(r)/r$.  Since\begin{equation} 
r\frac{d}{dr}\left(\frac{\phi_{l}(r)}{r}\right)= 
\frac{d\phi_{l}(r)}{dr}-\frac{\phi_{l}(r)}{r},\end{equation}we see 
that the operator $d/dr$ must be replaced by $(d/dr-1/r)$ in the 
coupled equations for $\phi_{l}(r)$.  This replacement has been 
performed below.

Finally, after noting that $\lambda$ is always even in this 
application, we have\begin{equation}\frac{\sqrt{4\pi}}{2\lambda+1}\langle 
lj\|T_{\pm}^{\lambda}\|l'j'\rangle= 
C^{\lambda}(j'jL_{\pm})\sum_{l_{\alpha}j_{\alpha}} 
A^{\lambda}(j_{\alpha}j'jL_{\pm}) \left[ \frac{d}{dr} +\frac{1}{2r} 
\left\{ l'(l'+1)-l_{\alpha}(l_{\alpha}+1)\right\} 
\right],\end{equation}where
 
\begin{equation}C^{\lambda}(j'jL_{\pm})= 
(-1)^{j'+\frac{1}{2}}\sqrt{\frac{36(2j'+1)(2L_{\pm}+1)(2j+1)(2l'+1)} 
{(2\lambda+1)}},\label{cj'j}\end{equation}and
\[A^{\lambda}(j_{\alpha}j'jL_{\pm}) = 
(2j_{\alpha}+1)\left \{\begin{array}{ccc}L_{\pm}&1&\lambda 
\\j'&j&j_{\alpha}\end{array}\right \}\left 
\{\begin{array}{ccc}l_{\alpha}&l'&1 \\
\frac{1}{2}&\frac{1}{2}&1 \\j_{\alpha}&j'&1 \end{array}\right \} 
\left(\begin{array}{ccc}j&L_{\pm}&j_{\alpha} 
\\\frac{1}{2}&0&-\frac{1}{2}\end{array} \right) \langle 
l'010|l_{\alpha}0\rangle.\]

After some rearrangement, we can write the reduced matrix element of 
the operator $\left[\nabla 
f_{\lambda\mu}(r)D^{\lambda}_{\mu0}({\bf\hat{r}})\right] 
\cdot(-i\nabla\times\hat{\sigma})$ as\begin{eqnarray}\lefteqn{\langle 
lj\|\left[\nabla 
f_{\lambda\mu}(r)D^{\lambda}_{\mu0}({\bf\hat{r}})\right] \!\cdot 
(-i\nabla\times\hat{\sigma}\|l'j'\rangle=}\nonumber\\
&&\left\{\displaystyle 
\sqrt{\lambda}Q_{\lambda\mu}(r)C^{\lambda}(j'jL_{-}) 
\sum_{l_{\alpha}j_{\alpha}} A^{\lambda}(j_{\alpha}j'jL_{-})- 
\sqrt{\lambda+1}P_{\lambda\mu}(r)C^{\lambda}(j'jL_{+}) 
\displaystyle\sum_{l_{\alpha}j_{\alpha}} 
A^{\lambda}(j_{\alpha}j'jL_{+})\right\}\frac{d}{dr}\nonumber\\
&&+\frac{1}{2r}\left\{\!\!\sqrt{\lambda} 
Q_{\lambda\mu}(r)C^{\lambda}(j'jL_{-}) 
\displaystyle\sum_{l_{\alpha}j_{\alpha}} 
A^{\lambda}(j_{\alpha}j'jL_{-}) 
\left[l'(l'+1)-l_{\alpha}(l_{\alpha}+1)\right]\right\}\nonumber\\ 
&&-\frac{1}{2r}\left\{\!\!\sqrt{\lambda+1} 
P_{\lambda\mu}(r)C^{\lambda}(j'jL_{+}) 
\displaystyle\sum_{l_{\alpha}j_{\alpha}} 
A^{\lambda}(j_{\alpha}j'jL_{+}) 
\left[l'('l'+1)-l_{\alpha}(l_{\alpha}+1)\right]\right\}\label{RMA}.
\end{eqnarray}

Recall that our original goal is to compute the spin-angular coupling 
matrix elements of the deformed spin-orbit interaction.  For 
$\lambda>0$ we have, from Eqs.  (\ref{soc}) and 
(\ref{WE}),\begin{equation}\langle 
lj\Omega|\delta V_{ls}(r,\theta,\phi)|l'j'\Omega'\rangle 
=2V_{so}\sum_{\lambda>0,\mu}\frac{\langle 
j'\Omega'\lambda\mu|j\Omega\rangle}{\sqrt{2j+1}}\langle 
lj\|\left[\nabla 
f_{\lambda\mu}(r)D^{\lambda}_{\mu0}({\bf\hat{r}})\right] \!\cdot 
(-i\nabla\times\hat{\sigma}\|l'j'\rangle\label{matel}.\end{equation}

\pagebreak


\begin{figure}
\caption{\label{vect}
(a) Relationship of the angular momentum vectors {\bf j}, {\bf R}, and 
{\bf I} in the axially-symmetric case.  (b) Same as (a) except for the 
triaxial case.  In both cases $K$ is the projection of the total 
angular momentum {\bf I} on the 3-axis of the body-centered system.}
\end{figure}
\begin{figure}
\caption{\label{width}
Calculated total decay width, in units of 10$^{-19}$ MeV, for 
$^{141g}$Ho, plotted as a function of the triaxial angle $\gamma$.  The 
shaded area represents the experimental measurement, and the error bar 
attached to the calculated curve represents the uncertainty in the 
calculated width due to the uncertainty in the proton energy.}
\end{figure}
\begin{figure}
\caption{\label{bratio}
Calculated branching ratio in percent for the decay of $^{141g}$Ho to 
the 2$^{+}$ state of $^{140}$Dy at an excitation energy of 202 keV, 
plotted as a function of the triaxial angle $\gamma$.  The shaded area 
represents the experimental measurement, and the error bar attached to 
the calculated curve represents the uncertainty in the calculated 
branching ratio due to the uncertainty in the proton energy.}
\end{figure}
\begin{figure}
\caption{\label{cj}
Amplitudes for various $\Omega$-values of the $j=7/2^{-}$ component of 
the $^{141g}$Ho wave function, plotted as a function of the triaxial 
angle $\gamma$.}
\end{figure}

\begin{table}
\caption{\label{jomega} Combinations of $j$ and $\Omega$ for 
$I=K=\frac{7}{2}^{-}$ originating from 
$j=\frac{7}{2}^{-}\rightarrow\frac{15}{2}^{-}$.  The bullet 
($\bullet$) denotes states occurring in the axially-symmetric case.}
\begin{ruledtabular}
\begin{tabular}{lrc|lrc}
$j$ & $\Omega$ & Axially-Symmetric & $j$ & $\Omega$ & Axially-Symmetric \\ \hline
$\frac{7}{2}^{-}$ & $-\frac{5}{2}$ & & $\frac{13}{2}^{-}$ & $-\frac{13}{2}$ \\
$\frac{7}{2}^{-}$ & $-\frac{1}{2}$ & & $\frac{13}{2}^{-}$ & $-\frac{9}{2}$\\
$\frac{7}{2}^{-}$ & $\frac{3}{2}$ & & $\frac{13}{2}^{-}$ & $-\frac{5}{2}$\\
$\frac{7}{2}^{-}$ & $\frac{7}{2}$ & $\bullet$ & $\frac{13}{2}^{-}$ & $-\frac{1}{2}$\\
$\frac{9}{2}^{-}$ & $-\frac{9}{2}$ & & $\frac{13}{2}^{-}$ & 
$\frac{3}{2}$\\
$\frac{9}{2}^{-}$ & $-\frac{5}{2}$ & &$\frac{13}{2}^{-}$ & $\frac{7}{2}$ & $\bullet$\\
$\frac{9}{2}^{-}$ & $-\frac{1}{2}$ & &$\frac{13}{2}^{-}$ & $\frac{11}{2}$\\
$\frac{9}{2}^{-}$ & $\frac{3}{2}$ & &$\frac{15}{2}^{-}$ & $-\frac{13}{2}$\\
$\frac{9}{2}^{-}$ & $\frac{7}{2}$ & $\bullet$& $\frac{15}{2}^{-}$ & 
$-\frac{9}{2}$\\
$\frac{11}{2}^{-}$ & $-\frac{9}{2}$ & &$\frac{15}{2}^{-}$ & 
$-\frac{5}{2}$\\
$\frac{11}{2}^{-}$ & $-\frac{5}{2}$ & &$\frac{15}{2}^{-}$ & $-\frac{1}{2}$\\
$\frac{11}{2}^{-}$ & $-\frac{1}{2}$ & &$\frac{15}{2}^{-}$ & $\frac{3}{2}$\\
$\frac{11}{2}^{-}$ & $\frac{3}{2}$ & &$\frac{15}{2}^{-}$ & 
$\frac{7}{2}$ & $\bullet$\\
$\frac{11}{2}^{-}$ & $\frac{7}{2}$ & $\bullet$ &$\frac{15}{2}^{-}$ & 
$\frac{11}{2}$\\
$\frac{11}{2}^{-}$ & $\frac{11}{2}$& &$\frac{15}{2}^{-}$ & $\frac{15}{2}$

\end{tabular}
\end{ruledtabular}
\end{table}

\begin{table}
\caption{\label{141} Comparison of calculated and experimental values 
for $^{141}$Ho(7/2$^{-}$) total decay rate and 2$^{+}$ branching 
ratio.  Input values are: $E_{x}(2^{+}$) in $^{140}$Dy = 202 keV, 
$\beta_{2} = 0.244$, and $\beta_{4} = -0.046$.  Calculations were done 
in the adiabatic limit , with no static triaxiality, and uncertainties 
coming from the measured proton energy of 1169(8) keV 
\protect\cite{davids1} and the decay half-life of 4.2(4) ms 
\protect\cite{davids1} have been taken into account.  The experimental 
spectroscopic factor $S_{exp}$ is defined as 
$\Gamma_{exp}/\Gamma_{calc}$.  The calculated spectroscopic factor 
$S_{calc}$ = $u^{2}$ is obtained from a BCS calculation (see text).}
\begin{ruledtabular}
\begin{tabular}{lccccc}
$\Gamma_{calc}$ (10$^{-19}$ MeV)& $\Gamma_{exp}$ 
(10$^{-19}$ MeV)& $S_{calc}$ & $S_{exp}$ &$BR(2^{+})_{calc}$ & 
$BR(2^{+})_{exp}$ \\ \hline
1.51$^{+.34}_{-.28}$ & 1.09(10)\footnote{ Ref.  \cite{davids1}} & 0.6 
& 0.72$^{+.17}_{-.15}$ &0.0071(5) & 0.0070(15)\footnote{ Ref.  
\cite{ryka}}
\end{tabular}
\end{ruledtabular}
\end{table}
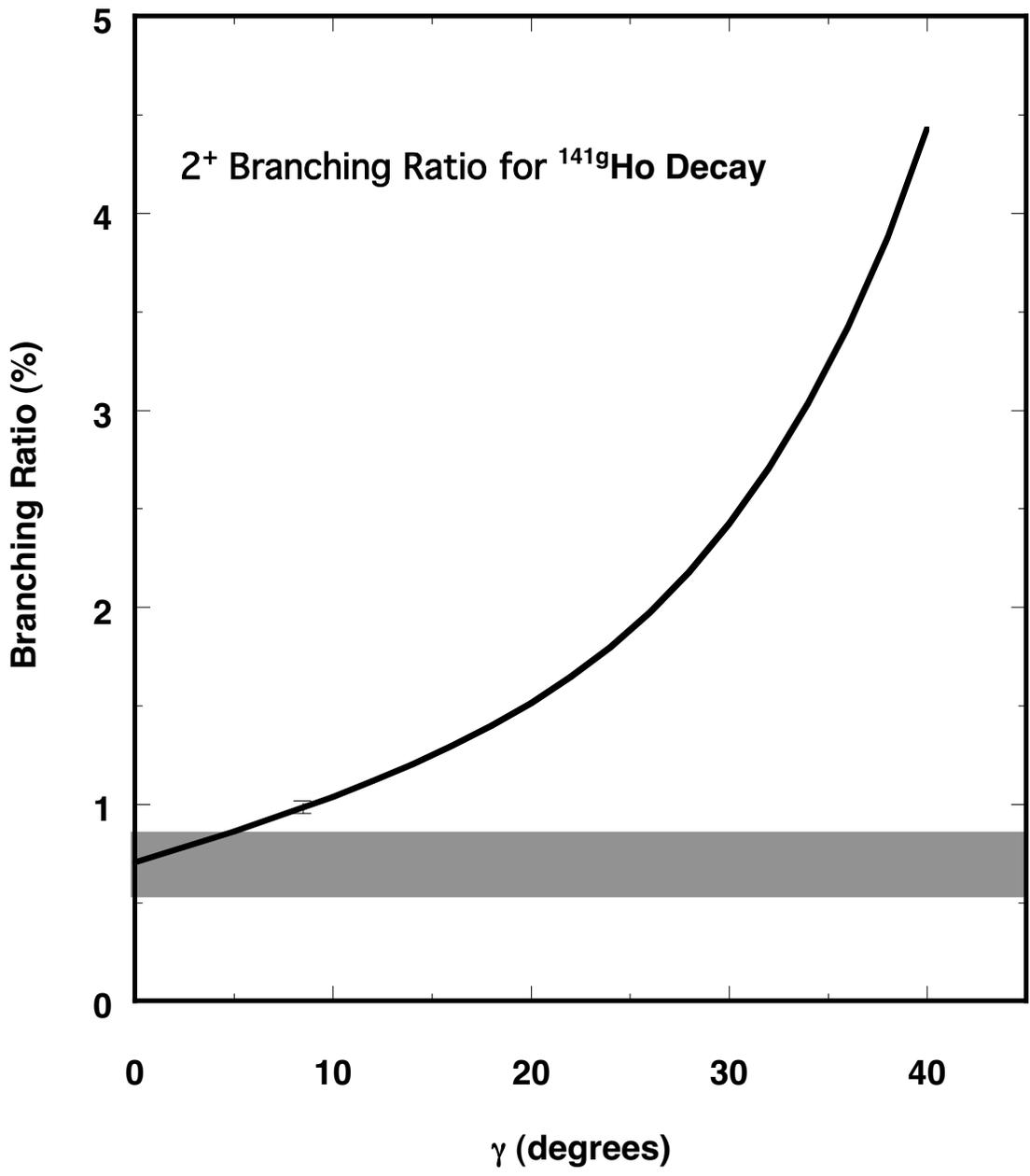
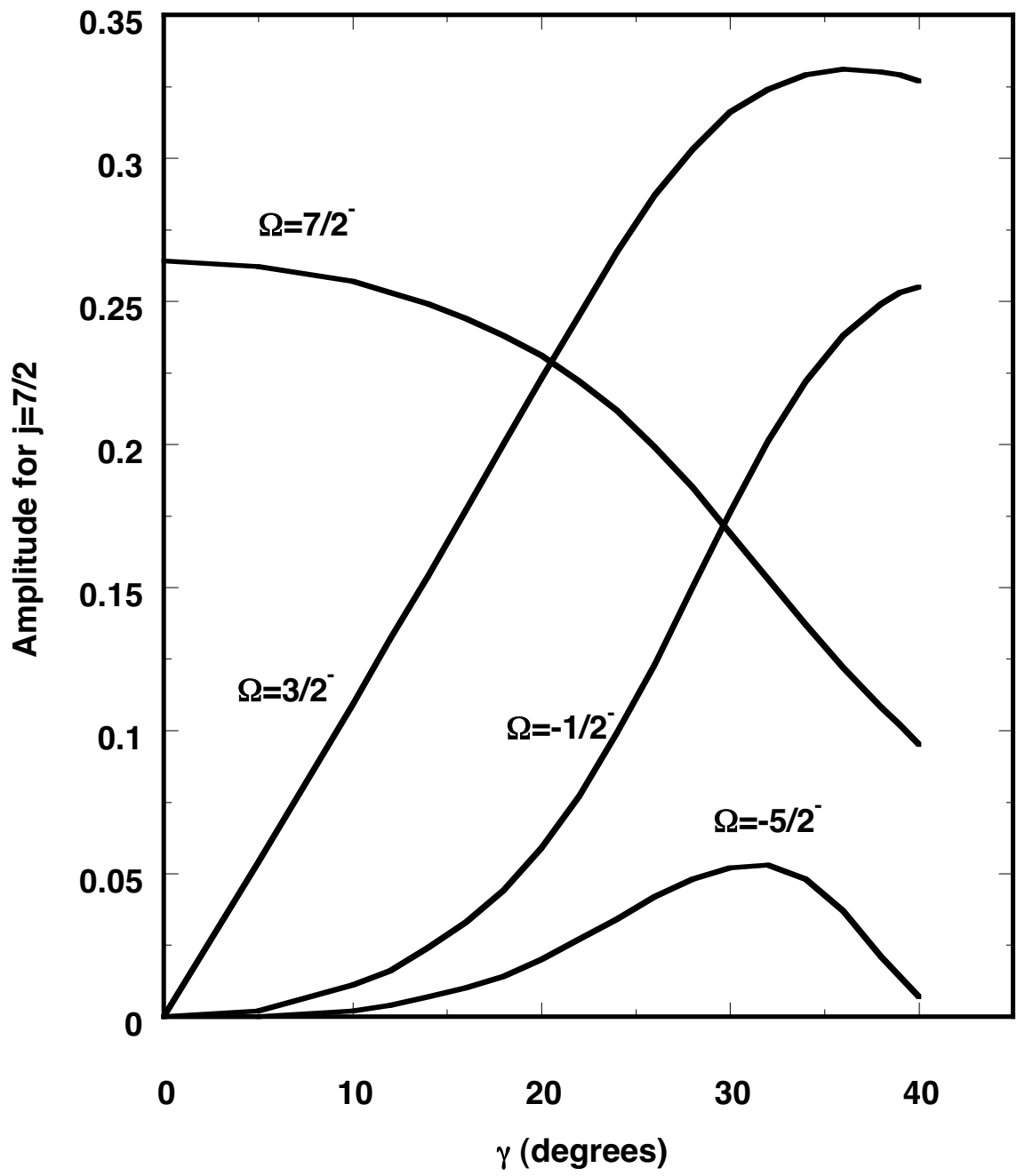
\end{document}